\documentclass[twocolumn]{aastex631}
\usepackage{amsmath}
\DeclareMathOperator{\mathion}{ion}
\usepackage{newfloat}
\DeclareFloatingEnvironment[name={Listing}]{listing}

\shorttitle{The \Lw{} Radiative Transfer Framework}
\shortauthors{Osborne \& Mili\'{c}}
\graphicspath{{./}{figures/}}

\begin{document}

\title{The \Lw{} Framework for NLTE Radiative Transfer in Python}

\correspondingauthor{Christopher M. J. Osborne}
\email{c.osborne.1@research.gla.ac.uk}

\author[0000-0002-2299-2800]{Christopher M. J. Osborne}
\affiliation{SUPA School of Physics and Astronomy,\\
University of Glasgow,\\
Glasgow, G12 8QQ, UK}

\author{Ivan Mili\'{c}}
\affiliation{Department of Physics,\\
University of Colorado,\\
Boulder CO 80309, USA}
\affiliation{Laboratory for Atmospheric and Space Physics,\\
University of Colorado,\\
Boulder CO 80303, USA}
\affiliation{National Solar Observatory,\\
Boulder CO 80303, USA}

\newcommand*{\Ca}{Ca\textsc{ii}}
\newcommand*{\CaLine}{Ca\,\textsc{ii} $8542$\,\AA}
\newcommand*{\Ha}{H$\alpha$}
\newcommand*{\Lw}{\textit{Lightweaver}}
\newcommand*{\MsLw}{\textit{MsLightweaver}}
\newcommand*{\NeedRef}{{\color{red} REF}}



\begin{abstract}

Tools for computing detailed optically thick spectral line profiles out of local thermodynamic equilibrium have always been focused on speed, due to the large computational effort involved.
With the \Lw{} framework, we have produced a more flexible, modular toolkit for building custom tools in a high-level language, Python, without sacrificing speed against the current state of the art.
The goal of providing a more flexible method for constructing these complex simulations is to decrease the barrier to entry and allow more rapid exploration of the field.

In this paper we present an overview of the theory of optically thick NLTE radiative transfer, the numerical methods implemented in \Lw{} including the problems of time-dependent populations and charge-conservation, as well as an overview of the components most users will interact with, to demonstrate their flexibility.

\end{abstract}

\keywords{Radiative Transfer (1335) ---
Radiative Transfer Simulations (1967) ---
Computational Methods (1965) ---
Solar Physics (1476) ---
Stellar Physics (1621)}


\section{Introduction}

Optically thick non-local-thermodynamic equilibrium (NLTE) radiative transfer is one of the most computationally intensive problems in modern solar and stellar physics.
It consists of taking a model atmosphere and computing self-consistent atomic populations whilst taking into account the fact that radiation originating from these atomic transitions may also affect their states elsewhere in the atmosphere.
The high numerical cost of this problem is due in part to the high dimensionality of the the intensity, as it varies with wavelength and direction in addition to the spatial and temporal variation of most other quantities considered, and also the possibly large number of contributors at each wavelength.
The NLTE problem can be extended to take into account the problem of finding an electron density consistent with the atomic populations, and this will also be discussed.

In recent times there has been a rise of flexible high performance frameworks available in high-level languages such as Python.
One domain where these have demonstrated their power is machine learning, where the building blocks provided by the frameworks allow researchers to rapidly prototype new systems with little loss in performance over a hand-tuned highly specific low-level implementation.
The goal of \Lw{} is to provide a similar set of tools for plane-parallel optically thick radiative transfer.
To this end it consists of an extensible Python frontend with a clean high-performance C++ backend.
During development the code has been extensively tested against both RH \citep{Uitenbroek2001,Pereira2015} and SNAPI \citep{Milic2018}, to ensure agreement between all three on a range of problems.
Whilst most radiative transfer tools are designed specifically for a single task, there is much commonality between the operations performed (especially the most costly operations, such as the formal solution of the radiative transfer equation).
It is therefore reasonable to abstract out these common building blocks in a way that allows a user to quickly build what amounts to a specialised tool with very little code, in a high-level, memory-safe language that is widely supported in the scientific computing community.

This report describes in detail the components of the \Lw{} framework, including the numerical methods used.
In Section \ref{Sec:Nlte} we provide an overview of NLTE radiative transfer and describe the numerical methods and their implementations.
Then in Section \ref{Sec:CodeDescription} the structure of the framework is described to demonstrate how modularity is achieved.

\Lw{} can be installed by an end-user through the standard Python package manager \texttt{pip} without need for particular compilers to be installed.
The code is freely available under the permissive MIT license\footnote{\url{https://opensource.org/licenses/MIT}} and is available on GitHub\footnote{\url{https://github.com/Goobley/Lightweaver}} with archival on Zenodo \citep{LightweaverZenodo}.
\Lw{} is in constant development and suggestions and enhancements are welcomed by contacting the authors or through the software's repository.
All examples in this paper were tested against the most recent release of \Lw{}, v0.5.0.
These examples are available on Zenodo \citep{LightweaverTrRepository}.

\section{Numerical NLTE Radiative Transfer}\label{Sec:Nlte}

In this Section we first present a brief overview of NLTE radiative transfer; for a \textit{much} more in depth introduction see \citet{Hubeny2014}.
We also explain how most terms are implemented in \Lw{}, especially those that are less apparent.

Solving the NLTE radiative transfer problem consists primarily of two coupled sub-problems:
\begin{itemize}
    \item Solving the radiative transfer equation to obtain the specific intensity at each frequency, point, and direction in the discretised computational domain for a given set of atomic populations and a given atmospheric model. This step is known as the formal solution of the radiative transfer equation.
    \item Updating the populations based on the radiative rates obtained from the formal solution.
\end{itemize}
These two problems are solved iteratively; a formal solution is first computed from an initial guess of the atomic populations, which is then used to construct a linear operator applied in conjunction with the current populations to update these until convergence.
Note that to compute the population update it is necessary to compute the intensity from transitions that do not overlap the spectral range of interest, due to their effect on the balance of transitions between the atomic population levels.

In the following we will expand on the solution of these two problems, first describing the terms that enter the equations, the construction of a linear operator, and the formal solution to the radiative transfer equation.

\subsection{Basic Definitions}\label{Sec:Basic}

The most basic quantity to consider in the study of radiation and radiation transport is specific intensity.
This is commonly denoted $I(\nu, \vec{d})$ at some frequency $\nu$ and direction $\vec{d}$ and has (SI) units J~m$^{-2}$~s$^{-1}$~Hz$^{-1}$~sr$^{-1}$.
Specific intensity (and its projections in the Stokes vector) is the quantity where our observations and simulations meet, and the only vector by which spectroscopic (and polarimetric) information arrives from the observed object.

A ray travelling through a medium, such as neutral gas or a plasma, gains a certain amount of energy per unit length due to emission processes in the plasma, and loses another amount due to absorption ($\chi_\mathrm{abs}$) and scattering ($\chi_\mathrm{scatt}$) processes.
These gain and loss terms are called emissivity and opacity, typically denoted $\eta$ and $\chi = \chi_\mathrm{abs} + \chi_\mathrm{scatt}$ respectively, and will depend both on the frequency considered, as well as the location and direction of the ray.

Considering the case of a ray travelling through plasma made up of neutral and ionised atoms, the emissivity and opacity will depend on the number of atoms, the frequency-dependent cross-sections of these atoms, and the quantum mechanical processes coupling the photons and the plasma.


For a bound-bound process we then arrive at the following expression for the radiative rates for a transition from level $i$ to level $j$ ($j > i$)
\begin{align}
    R_{ij} &= \oint \int B_{ij} \phi(\nu, \vec{d}) I(\nu, \vec{d})\,d\nu\,d\Omega,\\
    R_{ji} &= \oint \int \left[\left(A_{ji} + B_{ji} I(\nu, \vec{d})\right)\psi(\nu, \vec{d}) \right]\,d\nu\,d\Omega,
\end{align}
where $\phi$ is the line absorption profile, $\psi$ is the line emission profile, $A$ and $B$ are the Einstein coefficients for the transition.
The radiative rates have units s$^{-1}$
, and locally describe the number of atomic transitions ($i\rightarrow j$ and $j\rightarrow i$ respectively) per unit time.
For bound-free transitions the radiative rates are given by
\begin{align}
    R_{ij} &= \oint \int \alpha_{ij}(\nu) I(\nu, \vec{d})\,d\nu\,d\Omega,\\
    \begin{split}
    R_{ji} &= \oint \int \bigg[\left(I(\nu, \vec{d}) + \frac{2h\nu^3}{c^2}\right) \\
    &\hspace{2em}\cdot\alpha_{ij}(\nu)n_e\Phi_{ij}(T) e^{-h\nu/k_B T} \bigg]\,d\nu\,d\Omega,
    \end{split}
\end{align}
where $h$ is Planck's constant, $c$ is the speed of light, $\alpha_{ij}$ is the photoionisation cross-section, $k_B$ is Boltzmann's constant, and $n_e$ is the electron number density.
$\Phi$ is the Saha-Boltzmann equation defined such that

\begin{equation}
    \begin{split}
    n_e\Phi_{ij}(T) &= \frac{n^*_i}{n^*_j}\\
     &= \frac{g_i}{2g_j}\left( \frac{h^2}
    {2\pi m_e k_B T} \right)^{3/2} \exp{\left(  \frac{\Delta E_{ji}}{k_B T}\right)},
    \end{split}
\end{equation}
where $n^*$ is the population of the species in LTE, $m_e$ is the electron mass, $\Delta E_{ji}$ is the energy difference between levels $j$ and $i$, and $g_i$ is the statistical weight of level $i$. The Saha-Boltzmann equation is obtained by combining the Saha ionisation equation and the Boltzmann excitation equation, and describes the distribution of the total atomic population across its possible states, at a given electron density, and under the assumption of local thermodynamic equilibrium (LTE).

The rest frequency $\nu_{ij}$ of a transition is given by
\begin{equation}
    \nu_{ij} = \frac{h}{\Delta E_{ji}}.
\end{equation}
In \Lw{}, due to existing convention in other RT codes, the energy of each level relative to the ground level of the model atom is supplied in cm$^{-1}$.

The Einstein coefficients are related to each other and to the oscillator strength (a dimensionless quantity describing absorption probability) by
\begin{align}
    A_{ji} &= \frac{2\pi e^2 \nu_{ij}^2}{\epsilon_0 m_e c^3} f_{ij}, \\
    B_{ji} &= \frac{c^2}{2h\nu_{ij}^3} A_{ji}, \\
    B_{ij} &= \frac{g_j}{g_i} B_{ji},
\end{align}
where $f_{ij}$ is the oscillator strength, $e$ is the charge of an electron, and $\epsilon_0$ is the vacuum permittivity.
As the oscillator strength can be used to compute the Einstein coefficients for a transition, and again for consistency with other codes, \Lw{} requires $f_{ij}$ for each spectral line.

\subsection{Line Broadening}

A transition between well-defined energy states in a bulk motionless plasma is not infinitely narrow, but instead broadened by a number of factors, including natural radiative broadening, Doppler broadening, and collisional broadening, to give the absorption profile $\phi_{ij}$.
By default we follow the standard assumption of a Voigt absorption profile, and allow for a combination of different damping terms. There are described in detail in Appendix~\ref{App:LineBroadening}.

The design of \Lw{} also supports non-Voigt line profiles, such as the more complex model for electric pressure broadening discussed in \citet{Kowalski2017}, whilst only modifying the Python code in the model atom object, however the standard code-path for the Voigt profile is more optimised. The example presented in Section~\ref{Sec:AdvancedExample} shows how a Doppler line profile could be implemented.

\subsection{MALI}\label{Sec:Mali}

In the following we present a brief review of the numerical techniques implemented in \Lw{}.
Much of the content follows \citet{Uitenbroek2001} and the RH code described therein.

Given an atomic species where the population of excitation level $i$ is given by $n_i$ the general form of the kinetic equilibrium equation is given by

\begin{equation}
    \label{Eq:KinEq}
    \frac{\partial n_i}{\partial t} + \nabla \cdot (n_i \vec{v})
    = \sum_{j\neq i} n_j P_{ji} - n_i \sum_{j\neq i} P_{ij},
\end{equation}
where $\vec{v}$ is the bulk macroscopic velocity of the particle distribution, $P_{ij}$ is the total transition rate between atomic states $i$ and $j$ and is given by

\begin{equation}
    P_{ij} = R_{ij} + C_{ij},
\end{equation}
where $R_{ij}$ is the radiative rate, due to interaction with photons or spontaneous emission, and $C_{ij}$ is the collisional rate, due to interaction with other particles.
In NLTE studies it is normally assumed that the collisional rate can be known \textit{a priori} from the model atmosphere definition.

A common simplification of \eqref{Eq:KinEq} is to assume that the atmosphere is in a steady state (i.e. $\partial n_i / \partial t = 0$), and therefore the advective term can also be ignored.
This sets the left-hand side of \eqref{Eq:KinEq} to $0$ and we obtain the statistical equilibrium equation
\begin{equation}
    \label{Eq:StatEq}
    \sum_{j\neq i} n_j P_{ji} - n_i \sum_{j\neq i} P_{ij} = 0.
\end{equation}
For both \eqref{Eq:KinEq} and \eqref{Eq:StatEq}, the system must be solved simultaneously for all levels of an atomic species.
The latter requiring a constraint equation to avoid degeneracies.

\Lw{} adopts the same Rybicki-Hummer full preconditioning approach \citep{Rybicki1992} as used in RH \citep{Uitenbroek2001}, although the implementation is slightly different.
Following these authors, we write the emissivity $\eta$ and opacity $\chi$ of a transition between atomic levels $i$ and $j$, at frequency $\nu$, along a ray of direction $\vec{d}$ as

\begin{align}
    \label{Eq:Emis}
    \eta_{ij} &= n_j U_{ji}(\nu, \vec{d}), \\
    \label{Eq:Opac}
    \chi_{ij} &= n_i V_{ij}(\nu, \vec{d}) - n_j V_{ji}(\nu, \vec{d}),
\end{align}

where $n_i$ is the population density level $i$.
We assume here that $j > i$ and then, by convention, $\chi_{ji} = -\chi_{ij}$.

The $U$ and $V$ terms are defined for bound-bound and bound-free transitions as
\newlength{\WidestCase}
\settowidth{\WidestCase}{$n_e\Phi_{ij}(T)\left(\frac{2h\nu^3}{c^2}\right)e^{-h\nu/k_B T}\alpha_{ij}(\nu),$}
\begin{align}
    \label{Eq:Uji}
    U_{ji} =&
    \begin{cases}
        \frac{h\nu}{4\pi}A_{ji}\psi_{ij}(\nu, \vec{d}), & \textrm{bound-bound} \\
        n_e\Phi_{ij}(T)\left(\frac{2h\nu^3}{c^2}\right)e^{-h\nu/k_B T}\alpha_{ij}(\nu), & \textrm{bound-free},
    \end{cases}\\
    \label{Eq:Vij}
    V_{ij} =&
    \begin{cases}
        \makebox[\WidestCase][l]{$\frac{h\nu}{4\pi}B_{ij}\phi_{ij}(\nu, \vec{d}),$} & \textrm{bound-bound} \\
        n_e\Phi_{ij}(T)e^{-h\nu/k_B T}\alpha_{ij}(\nu), & \textrm{bound-free},
    \end{cases}\\
    \label{Eq:Vji}
    V_{ji} =&
    \begin{cases}
        \makebox[\WidestCase][l]{$\frac{h\nu}{4\pi}B_{ji}\psi_{ij}(\nu, \vec{d}),$} & \textrm{bound-bound} \\
        \alpha_{ij}(\nu), & \textrm{bound-free}.
    \end{cases}
\end{align}
By convention we define $U_{ij} = U_{ii} = V_{ii} = 0$.

\Lw{} can also treat lines necessitating partial redistribution (PRD). Following \citet{Uitenbroek2001} we define

\begin{equation}
    \label{Eq:RhoDef}
    \rho_{ij}(\nu, \vec{d}) = \frac{\psi_{ij}(\nu, \vec{d})}{\phi_{ij}(\nu, \vec{d})}
\end{equation}

and thus

\begin{align}
    \label{Eq:UjiRho}
    U_{ji} &= \frac{h\nu}{4\pi}A_{ji}\rho_{ij}(\nu, \vec{d})\phi_{ij}(\nu, \vec{d}), \quad \textrm{bound-bound}\\
    \label{Eq:VjiRho}
    V_{ji} &= \frac{h\nu}{4\pi}B_{ji}\rho_{ij}(\nu, \vec{d})\phi_{ij}(\nu, \vec{d}), \quad \textrm{bound-bound}.
\end{align}
In the case of complete redistribution (CRD) $\rho=1$.
These terms will be discussed in detail in Section~\ref{Sec:Prd}.

The total opacity and emissivity can then be found by summing over all species

\begin{align}
    \label{Eq:TotEmis}
    \eta_{\mathrm{tot}}(\nu, \vec{d}) &= \sum_{\mathrm{species}} \left[ \sum_j \sum_{i<j} \eta_{ij}(\nu, \vec{d}) \right], \\
    \label{Eq:TotOpac}
    \chi_{\mathrm{tot}}(\nu, \vec{d}) &= \sum_{\mathrm{species}} \left[ \sum_j \sum_{i<j} \chi_{ij}(\nu, \vec{d}) \right].
\end{align}

In \Lw{} we split species into three categories:
\begin{itemize}
    \item \textbf{Background:} Bound-free transitions are considered under the assumption of LTE. The opacity and emissivity contribution here is considered to be isotropic.
    \item \textbf{Detailed:} All transitions (bound-bound and bound-free) are considered in detail, using either given (e.g. from a previous NLTE simulation) or LTE populations. The opacity and emissivity contribution here is considered to be angle-dependent.
    \item \textbf{Active:} All transitions are considered in detail and terms necessary for iterating the populations are accumulated. The opacity and emissivity contribution here is considered to be angle-dependent.
\end{itemize}

The expressions for total emissivity and opacity can then be written as the summation over the emissivity and opacity in each of the three previous categories, for each frequency and direction.

The source function at a given frequency and direction is then given by
\begin{equation}
    S(\nu, \vec{d}) = \frac{\eta_\mathrm{tot}(\nu, \vec{d}) + \sigma(\nu) J(\nu)}{\chi_\mathrm{tot}(\nu, \vec{d})},
\end{equation}
where $\sigma$ is the continuum scattering coefficient that will be discussed further in Section \ref{Sec:Background}.

It is common to define an operator, $\Lambda$ used to obtain the monochromatic radiation field in a particular direction from the source function

\begin{equation}
    I(\nu, \vec{d}) = \Lambda_{\nu, \vec{d}}\left[S(\nu, \vec{d})\right].
\end{equation}
In essence, this is our formal solver, discussed in Section \ref{Sec:Fs}.
\citet{Rybicki1992} introduce an additional operator, $\Psi$, that aids in the construction of a linear preconditioned iterative scheme for solving \eqref{Eq:StatEq} such that

\begin{equation}
    \Psi_{\nu, \vec{d}}\left[\eta_\mathrm{tot}(\nu, \vec{d})\right] =
    \Lambda_{\nu, \vec{d}}\left[\frac{\eta_\mathrm{tot}(\nu, \vec{d})}{\chi^\dagger_\mathrm{tot}(\nu, \vec{d})}\right],
\end{equation}

where $\chi^\dagger_\mathrm{tot}$ is the opacity evaluated with the populations from the previous iteration.
For the converged solution, these two operators are equivalent, as $\chi^\dagger=\chi$.

Taking the $\vec{n}$ as the vector of level populations $\{n_1, n_2, \ldots, n_N\}$ at a location in the atmosphere, we can write our iterative scheme for \eqref{Eq:KinEq} as
\begin{equation}
    \label{Eq:GammaEse}
    \frac{\partial n_i}{\partial t} + \nabla \cdot (n_i \vec{v}) = \Gamma_i \vec{n},
\end{equation}
where $\Gamma_i$ is a row vector from the matrix $\Gamma = \Gamma^C + \Gamma^R$, which is evaluated using the previous population estimate. $\Gamma^C$ and $\Gamma^R$ represent the preconditioned collisional and radiative rate equations respectively.
We will address the construction of $\Gamma^C$ later.
From \citet{Rybicki1992} and \citet{Uitenbroek2001} we can write
\begin{align}
\begin{split}\label{Eq:GammaR}
    \Gamma^R_{ll^\prime} = \oint \int \frac{1}{h\nu} \bigg( &U^\dagger_{l^\prime l} + V^\dagger_{l^\prime l}I_{\nu, \vec{d}}^\mathrm{eff} - \\
    &\left(\sum_{m\neq l}\chi^\dagger_{lm}\right) \Psi^*_{\nu, \vec{d}} \left[ \sum_p U^\dagger_{l^\prime p} \right] \bigg)\, d\nu\,d\Omega
\end{split}
\end{align}
for $l\neq l^\prime$, and all $\dagger$ terms evaluated with the current level population and $\rho$ estimates.
The term
\begin{equation}\label{Eq:Ieff}
    I^\mathrm{eff}_{\nu, \vec{d}} = I^\dagger(\nu, \vec{d}) - \Psi^*_{\nu, \vec{d}}\left[ \sum_{i,j} \eta^\dagger_{ij} \right],
\end{equation}
when assuming a diagonal $\Psi^*$ operator, describes the non-local contribution to the radiation field from the atom in question, and the local contribution from other species.
It is often separated, as it remains constant for all transitions in an atom.
The diagonal terms of $\Gamma$ are computed using the conservation property that requires, for the sake of total number conservation, that the sum of each column of $\Gamma$ be zero \citep{Rybicki1992}.
Thus,
\begin{equation}
    \Gamma_{ll} = -\sum_{m\neq l} \Gamma_{ml}.
\end{equation}

Now that $\Gamma$ has been constructed it can be used in \eqref{Eq:GammaEse}.
In the case of statistical equilibrium, we must solve the matrix-vector equation $\Gamma \vec{n} = \vec{0}$.
A constraint equation is also needed, to avoid the trivial solution ($\vec{n}=\vec{0}$), typically a constraint on the total number density of the species.
In effect, this amounts to replacing one of the equations with a sum over the level populations, i.e. replacing one of the \textit{rows} of $\Gamma$ with ones, and the associated entry in right-hand side with the total population number density.

The discretisation of the time-dependent form of the kinetic equilibrium equation is discussed in the following section as it involves extra complexities strongly coupled to the numerical methods applied.

\subsubsection{Numerical Implementation}\label{Sec:MaliImpl}
Most of the integration terms proceed similarly to those of the RH code, but as those have not been presented in a single document, we describe the numerical implementation in detail here.

When considering a one-dimensional plane parallel atmosphere, as is done in \Lw{}, it is efficient to discretise the integration over solid-angle using Gauss-Legendre quadrature over the cosine of the angle between the ray and the normal to atmospheric slabs, commonly denoted $\mu$. These integrations are then implemented as weighted summations of the integrand at the Gauss-Legendre nodes i.e. the angle averaged intensity
\begin{equation}
    J(\nu) = \frac{1}{4\pi}\oint I(\nu, \vec{d})\, d\Omega
\end{equation}
at a point in the atmosphere can be calculated from
\begin{equation}
    J(\nu) = \sum_\mu I(\nu, \vec{d}) w_\mu.
\end{equation}
The number of angle points is user defined, as it depends on the problem (the anisotropy of the radiation field): for static atmospheres three angle samplings are normally sufficient, whereas five is more reliable in dynamic atmospheres.

As \Lw{} handles overlapping transitions, there needs to be a common wavelength grid that covers all transitions for the problem in question.
Each transition provides a set of wavelengths that need to be taken into account to reliably solve the RT problem (e.g. lines are typically densely sampled in the line core and sparse in the wings).
All of these individual wavelength grids are combined to produce the global wavelength grid, and a new grid is created for each transition which contains all of the original points, as well as the wavelength points from all other transitions that overlap.

These wavelength grids also define the basis of a numerical quadrature that is described in Appendix~\ref{App:MaliImpl}. Therein we also describe the specific accumulation terms used in the construction of the fully preconditioned $\Gamma$.

In the case of the time-dependent kinetic equilibrium, there is no ``one-size-fits-all`` approach to this equation and \Lw{} provides the following tools.
The advective term in \eqref{Eq:GammaEse} is ignored, as this requires a more complete treatment including consideration of hydrodynamics.
We can discretise $\partial \vec{n} / \partial t = \Gamma \vec{n}$ using a theta method
\begin{equation}
    \label{Eq:ThetaDisc}
    \frac{\vec{n}^{t+1} - \vec{n}^t}{\Delta t} = \theta \Gamma^{t+1} \vec{n}^{t+1} + (1-\theta)\Gamma^{t} \vec{n}^{t},
\end{equation}
where the superscripts $t$ and $t+1$ indicate the start and end of the timestep being integrated over, $\Delta t$ the duration of the timestep, and $\theta$ the degree of implicitness.
$\theta=0.5$ represents the Crank-Nicolson scheme, $\theta=1$ the backwards Euler scheme, and $\theta=0.55$ is commonly used as it is often found to cope better with stiff systems \citep[eg.][]{Viallet2011}.
This system is solved by storing $\Gamma^{t}$ at the start the process, and then updating $\Gamma^{t+1}$ using revised updates of the populations $\vec{n}^{t+1}$ with each iteration.
The process of obtaining a new estimate for $\vec{n}^{t+1}$ can be found by rearranging \eqref{Eq:ThetaDisc} into the form
\begin{equation}
    \label{Eq:TimeDepSystem}
    (\mathbb{I} - \theta\Delta t \Gamma^{t+1}) \vec{n}^{t+1} = (1-\theta)\Delta t \Gamma^{t}\vec{n}^{t} + \vec{n}^{t},
\end{equation}
where $\mathbb{I}$ is the identity matrix.
As the right-hand side is known \textit{a priori}, it can be evaluated directly, and \eqref{Eq:TimeDepSystem} is a matrix-vector system that can be solved equivalently to the statistical equilibrium case, albeit without the need for a constraint equation.
Currently only the fully implicit $\theta = 1$ case is supported, as during testing the differences were found to be insignificant, however we plan to include support for other $\theta$ in the future, and this has already been implemented in separate packages that use \Lw{}, but without modifying the base framework.

\subsection{Radiative Transfer Equation}

To obtain the intensity terms in the radiative rates, as well as the outgoing intensity, we need to solve the monochromatic radiative transfer equation (RTE), which for a one-dimensional plane-parallel atmosphere stratified along the $z$-axis is expressed as

\begin{equation}
    \label{Eq:Rte}
    \mu\frac{\partial I(\nu, \vec{d})}{\partial z} = \eta(\nu, \vec{d}) - \chi(\nu, \vec{d})I(\nu, \vec{d}),
\end{equation}
or along an optical-depth stratification as
\begin{equation}
    \label{Eq:RteTau}
    \mu\frac{\partial I(\nu, \vec{d})}{\partial \tau(\nu)} = I(\nu, \vec{d}) - S(\nu, \vec{d}),
\end{equation}
for the optical depth defined as $d \tau(\nu) = -\chi(\nu)\, dz$.

Solving this equation for multiple projected angles $\mu$ provides the radiation field throughout the model atmosphere that is necessary to compute the $\Gamma$ operator. Typically the optical depth formulation of \eqref{Eq:RteTau} is solved as it is more numerically robust \citep{Janett2018, DelaCruzRodriguez2013}.

\subsubsection{Formal Solver}\label{Sec:Fs}

The formal solver is the technique by which the RTE \eqref{Eq:Rte} is solved and the approximate operator $\Psi^*$ is computed.
By default we adopt the third order Bézier spline short-characteristics approach of \citet{DelaCruzRodriguez2013, 2019dlcr}, however investigation is also under way into the use of pragmatic formal solvers as discussed in \citet{Janett2018} and the BESSER formal solver of \citet{Stepan2013}.

In the short-characteristics approach the formal solver is provided with the opacity and source function at discretised points throughout the atmosphere, and the behaviour of these between the known points is assumed to follow a simple function that can be analytically integrated, in this case a third order Bézier spline.
It is important to choose an interpolating function that varies smoothly and minimises, or better yet, eliminates under- and overshoots in the interpolant.
The third order Bézier spline has proven to be robust in this setting and has been applied in other modern codes such as STiC \citep{2019dlcr} and SNAPI \citep{Milic2018}.

The integration routine proceeds from one end of the atmosphere to the other, accumulating these terms through the analytic short-characteristics integration to obtain the up- or down-going intensity for this ray at each point in the atmosphere.

The approximate $\Psi$ operator $\Psi^*$ is simply the diagonal of the true $\Psi$ operator, a matrix that would map the vector of emissivity to the intensity. $\Psi^*$ is trivially computed during the formal solution from the local contribution terms to the intensity and the local opacity.

The simple linear short-characteristics formal solver is also present and new formal solvers that conform to the interface used in \Lw{} can be compiled separately and loaded from a shared code library, allowing \Lw{} to serve as a testbed without need to modify the core package.

\subsection{Partial Frequency Redistribution}\label{Sec:Prd}

The effects of partial frequency redistribution (PRD) are important for some NLTE lines, typically strong resonance lines and lower density regions where radiative effects dominate over collisional effects \citep{Hubeny2014}.
For a complete treatment of the theory describing PRD lines we direct readers to \citet{Hubeny2014} and references therein, but we will provide a basic overview here.
The common assumption of complete frequency redistribution (CRD) in spectral lines is that $\psi=\phi$.
The argument is that most lines are formed in regions with sufficient elastic collisions that atoms are well distributed across the sub-states of each energy level.
Emission is therefore not correlated with the absorbed photon that excited the atom into this state.
When the plasma is less collisional, there is said to be a natural population of a particular level, i.e. a population where the emission frequency is correlated to the absorption frequency.
In this case the emission profile $\psi$ differs from the absorption profile $\phi$, and these coherent scattering effects must be considered.

\Lw{} currently adopts the iterative PRD approach presented in \citet{Uitenbroek2001}, but may also in future implement a direct solution, as it may prove more robust than the iterative approach for some highly dynamic problems, despite the higher computational cost.
Currently cross-redistribution is not implemented, but the ground-work is present, and the remaining changes would be a simple extension following \citet{Uitenbroek2001} and the RH code.

In the common case where flows are lower than the thermal Doppler velocity, the integrations needed to solve the PRD equations can be simplified by assuming isotropy of the radiation field. This is known as angle-averaged PRD. In cases with stronger flows we instead employ the hybrid PRD approach of \citet{Leenaarts2012} which consists of computing $\rho$ in the atom's rest frame. This approximation agrees quite well with a full angle-dependent treatment, is simple to implement, and much faster to evaluate than the full angle-dependent case.
Due to the additional computation effort involved in PRD calculations, regardless of the method used, lines need to be explicitly labelled as PRD.

The derivation of the PRD equations and their numerical implementation is described in Appendix~\ref{App:CompletePrdImpl}.

\subsection{Self-consistent Electron Density}\label{Sec:EleDens}

The MALI technique assumes that the electron density is known \textit{a priori}, but this is often not the case.
Assuming that the electron density can be given by the LTE ionisation state of the plasma can yield substantially incorrect results for chromospheric and prominence lines \citep{Heinzel1995,Paletou1995,Bjorgen2019}.
An additional iteration process is therefore needed to determine the correct electron density within the framework of the NLTE problem.

Whilst not quite as robust as the pure MALI treatment a secondary Newton-Raphson iteration to self-consistently compute electron density was proposed by \citet{Heinzel1995} and \citet{Paletou1995}, and forms the basis of the method implemented here.
The time-dependent case is based on \citet{Kasparova2003}, and the numerical implementation of both of these is described in Appendix~\ref{App:EleDens}.

\subsection{Collisional Rates}

Based on the RH code, a number of different formulations for collisional rates are available in \Lw{}.
Currently these include tabulated collision strength ($\Omega$) against temperature for excitation of ions by electrons, tabulated collisional ionisation and excitation rates of neutrals by electrons (known in RH as CI and CE), tabulated collisional excitation by protons, neutral hydrogen, and charge exchange with these species (CP, CH, CH+, and CH0 respectively).
Additionally the collisional ionisation rates of \citet{Arnaud1985}, and \citet{Burgess1983} are present.
These can be extended further in user code with no modifications to the base library.
The choice of pre-implemented collisional rates in the \Lw{} ``standard library'' allow the direct conversion of the majority of model atoms that are distributed with RH to also be distributed with \Lw{}.
The collisional rates for a level depend only on the local parameters, and have no wavelength dependence, therefore the implementation is much more straightforward and does not require complicated numerical integration.

By default, the collisional rates are re-evaluated at the start of each formal solution, although this can be disabled by the user.

\subsection{Full Stokes Treatment}

We also support Zeeman splitting and polarisation effects where the complete set of anomalous Zeeman splitting terms can be computed from the quantum numbers $J$, $L$, and $S$ for the levels considered through the $LS$ coupling formalism, or a classical Zeeman triplet computed from an effective Landé $g$-factor present in the definition of a line.
\Lw{} does not support full Stokes iteration of the populations, but provides support for both the field-free and polarisation-free approaches \citep{TrujilloBueno1996}.
The final formal solution is then undertaken with the third-order Bézier spline Diagonal Element Lambda Operator (DELO) method of \citet{DelaCruzRodriguez2013}.

\subsection{Miscellaneous}

Like RH, \Lw{} utilises base SI units throughout, with the singular exception of wavelength being treated in nm.
The units of a variable are therefore easy to determine, with little consideration of derived units.
In the remainder of this Section we will discuss other small implementation details of the code.

\subsubsection{Collisional-Radiative Switching}

The collisional-radiative switching (CRSW) technique of \citet{Hummer1988} is available in \Lw{}.
Using MALI, many problems will converge without much issue, however in the case of strong atmospheric gradients the corrections to the populations in early iterations can be overly large and drive the system into a poorly conditioned state.
To avoid this the CRSW technique multiplies the collisional contributions to $\Gamma$ by a significant factor, so as to force the system into LTE.
This factor is slowly reduced, allowing a graceful departure from LTE towards  NLTE.
The exact decay of this parameter can be configured by the user.

\subsubsection{Isotopes}

Isotopic models are also supported as valid atomic models.
By default the abundances for all elements and their isotopic proportions are taken from \citet{Asplund2009}, however these can easily be modified by the user.

\subsubsection{Equation of State}\label{Sec:Eos}

\Lw{} contains a simple equation of state and background opacity package based on \citet{1970stat.book.....M}, implemented by Wittmann, and ported to Python by J. de la Cruz Rodriguez\footnote{\url{https://github.com/jaimedelacruz/witt/}}.
This equation of state has also been used in SIR \citep{1992RuizCobo} and NICOLE \citep{Socas-Navarro2015}.
In \Lw{} it is often used to determine the values of unknown parameters in a provided model atmosphere, and determining an LTE hydrostatic stratification if necessary (based on NICOLE).
The equation of state also provides an estimate of the reference opacity $\tau_{500}$ at $500\, nm$ for model atmospheres that provide a height or column mass based stratification.

\subsubsection{Molecules}\label{Sec:Molecules}

Whilst molecular lines are not currently supported by \Lw{}, it can compute molecular formation in instantaneous chemical equilibrium, using the same molecular models as RH.
These molecules reduce the populations of the atoms bound up in them and some (OH, CH, and H$^-$) contribute to the background opacity.
The H$^-$ population is always computed, due to its importance in obtaining correct background opacities.

\subsubsection{Background Treatment}\label{Sec:Background}

The default implementation of background emissivities, opacities, and scattering terms currently follows that of RH, but a more general interface that is trivially overrideable in user code without modifying the framework is also present.
The components present in the default background opacity package are listed in Table~\ref{Tab:BgRefs}.
The OH and CH opacities are not present unless these molecules are explicitly loaded and instantaneous chemical equilibrium is computed as discussed in Section~\ref{Sec:Molecules}.

\begin{table*}
\centering
\begin{tabular}{ll}
\hline
Component                                   & Reference                       \\ \hline
H free-free                                 & \citet{Mihalas1978}             \\
H$_2^-$ free-free                           & \citet{Bell1980}                \\
H$_2^+$ free-free                           & \citet{Bates1952}               \\
H$_2$ Rayleigh scattering                   & \citet{Victor1969}              \\
H$^-$ bound-free                            & \citet{Geltman1962,Mihalas1978} \\
H$^-$ free-free                             & \citet{Stilley1970,Mihalas1978} \\
H$^-$ free-free ($>9113\,nm$) & \citet{John1988}                \\
OH bound-free                               & \citet{Kurucz1987}              \\
CH bound-free                               & \citet{Kurucz1987}              \\
General Rayleigh scattering                 & \citet{Mihalas1978}             \\ \hline
\end{tabular}
\caption{References for components present in default background opacity package.}
\label{Tab:BgRefs}
\end{table*}

\subsubsection{Interpolation}

For interpolation duties, other than those in the formal solver and calculation of the PRD terms, we adopt the rapid, but robust fourth-order weighted essentially non-oscillatory (WENO) approach presented in \citet{Janett2019}.
Whilst this technique does not guarantee monotonicity around discontinuities, the over- and under-shoots remain very small, with no ringing artifacts, and we feel that the high quality of the solution in smooth regions makes it worthwhile.
We have provided a performant implementation of this technique as a separate Python package that is available through \texttt{pip} as \texttt{weno4}, on GitHub\footnote{\url{https://github.com/Goobley/Weno4Interpolation}} and archived on Zenodo \citep{Weno4Release}.

\section{Description of Major Code Components}\label{Sec:CodeDescription}

In this section we provide a brief overview of the components of \Lw{} a user will typically interact with.
The frontend is entirely constructed in Python with a binding layer written in Cython\footnote{\url{https://cython.org/}} \citep{Behnel2011} to allow it communicate with the C++ backend.

\subsection{Atomic Models}\label{Sec:AtomicModels}

The information stored in model atoms used by contemporary codes is more than simple atomic data, and in essence these codes are defining their own \textit{ad hoc} scripting languages to support reading the various terms encoded in these files.
As we have access to a high-level dynamic language in the form of Python, it is reasonable to model these as object hierarchies where we can take advantage of the common Python convention that \texttt{obj == eval(repr(obj))} i.e. evaluating the textual representation of the object generates an equivalent object.

\begin{listing*}
\centering
\includegraphics{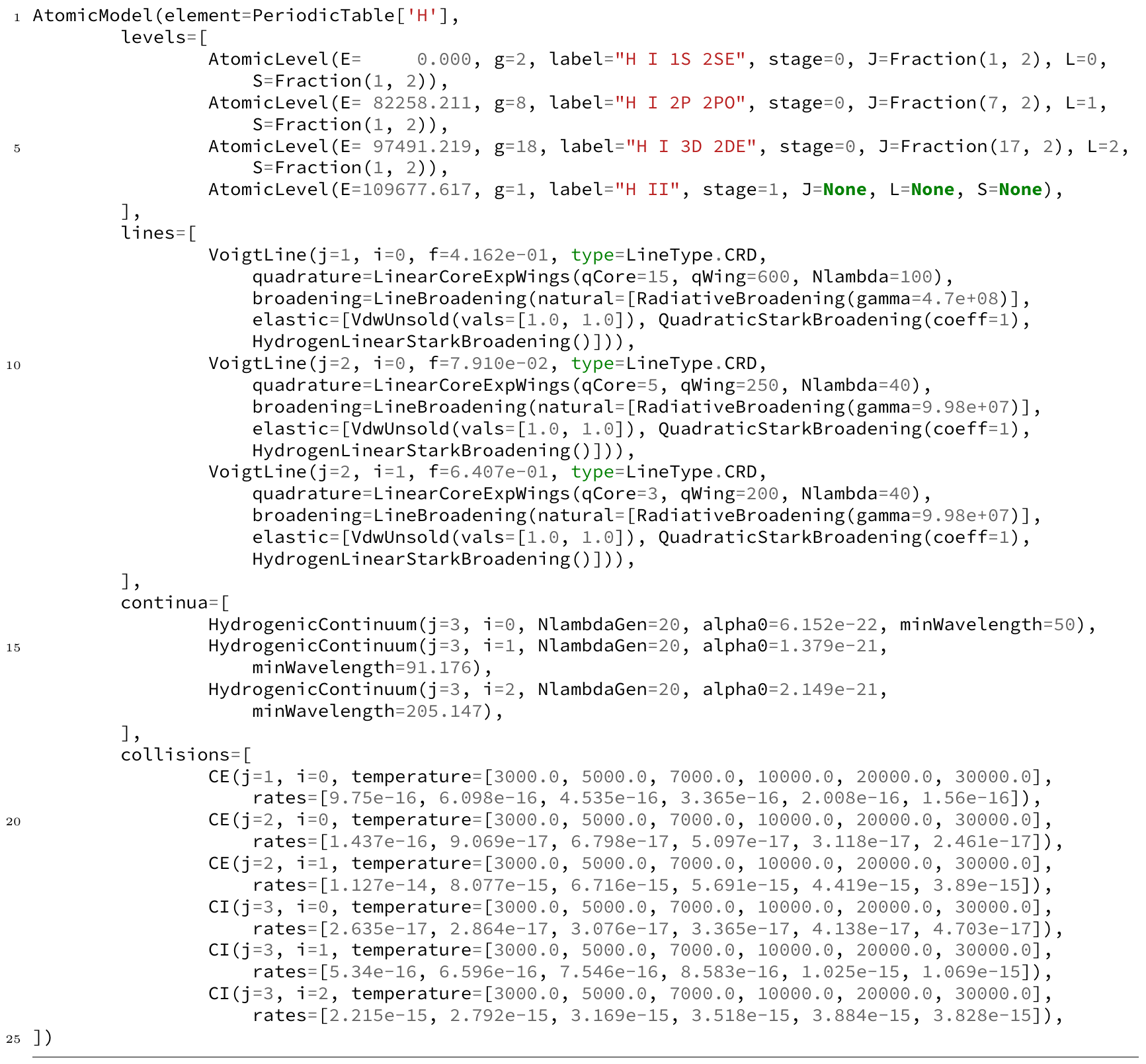}
\caption{Complete code for a 3 level + continuum H atomic model.}
\label{Lst:H4}
\end{listing*}

The code in Listing \ref{Lst:H4} shows the complete source necessary for a 3 level with continuum Hydrogen atom.
It is constructed from nested Python classes, and through inheritance user code that implements the same interfaces will be able to extend these further.

For example, taking the \texttt{quadrature} component of a spectral line, we see here that the \texttt{LinearCoreExpWings} class is used.
This is a derived class of \texttt{LineQuadrature} and any derived instance of this class can be used here.
The requirements are that it provide at least functions \texttt{doppler\_units}, \texttt{wavelength}, and \texttt{\_\_repr\_\_}, that return the quadrature in Doppler units and wavelength respectively, and specify how to print the object so it can be re-evaluated.
The last of these is trivial and there are plenty of examples throughout the \Lw{} codebase.
Line broadening terms are implemented similarly.
One strength of the model atoms being implemented in terms of objects is the ease at which they can be manipulated with a simple script before being used or saved in text form (or an optimised Python object storage format such as \texttt{pickle}\footnote{\url{https://docs.python.org/3/library/pickle.html}}).

\subsection{Radiative Set}\label{Sec:RadSet}

During the configuration of a simulation, all atomic models, whether ``active'' (full NLTE), ``detailed static'' (transitions computed in detail, but populations fixed), or ``passive'' (background contributions only) are stored in a \texttt{RadiativeSet} object.
This is responsible for producing the common wavelength grid from all transitions taken into account, and the final grid for each transition whilst taking the other transitions into account as discussed in Section \ref{Sec:MaliImpl}.
This data is returned in a \texttt{SpectrumConfiguration} object, that can create another instance of itself, covering a restricted range of wavelengths, that is often used for computing a final formal solution over a line in detail, after the NLTE iteration is complete.

The \texttt{RadiativeSet} is also responsible for determining the LTE populations of these species from their models and the atmospheric data provided.
During this process the electron density can be assumed fixed, as provided in the atmospheric data, or can be iterated to be self-consistent with the LTE populations.
In the future this object will also be responsible for optionally applying a variant of the second order escape probability method of \citet{Hummer1982}, applied to MALI by \citet{Judge2017}, which currently resides in the C++ backend.
This LTE and initial condition atomic population data is returned in a \texttt{SpeciesStateTable}.

\subsection{Species State Table}

The \texttt{SpeciesStateTable} is responsible for holding both the LTE and NLTE populations of the species (and molecules) present in the simulation, as well as the radiative rates for species treated in detail.
This object can also update the LTE and H$^-$ populations given an updated set of atmospheric data, thus facilitating time-dependent simulations.
The arrays in this object are updated automatically by the C++ backend, as they are in fact shared \textit{by reference}, and the backend is operating directly on the same memory, with no duplication necessary.
This is achieved with a lightweight C++ library allowing multi-dimensional views onto a non-owned segment of data.
These arrays provide a subset of NumPy functionality, and are limited to handling contiguous memory for performance.
Thanks to the use of C++ templates for various data types, these have been verified to compile to assembly equivalent to access into a flat array, with no performance loss, but substantially greater memory safety than raw pointers, and the option to enable bounds-checking during debugging (by adjusting compilation flags).

\subsection{Context}

The code objects discussed so far primarily describe the configuration of the simulation which is then controlled by the \texttt{Context} object.
The \texttt{Context} takes this data, in addition to several other configuration options, such as the whether to use hybrid PRD, charge conservation, CRSW, Ng acceleration \citep{Ng1974}, multi-threading options, and initial solution to use (which as discussed in Section \ref{Sec:RadSet}, will be moved to the frontend in future).
The effects of most of these options can also be achieved by calling some extra methods, but are simplified when used as arguments to the \texttt{Context} initialiser.
During initialisation the \texttt{Context} computes background opacity, emissivity and scattering, line profiles, and maps the data into a form which can be used by the backend.

After this initial setup the \texttt{Context} can be used to interact with the backend by calling various methods.
These include
\begin{itemize}
    \item \texttt{formal\_sol\_gamma\_matrices} which evaluates the collisional rates, formal solution for all wavelengths, and constructs the $\Gamma$ operator.
    \item \texttt{single\_stokes\_fs} which computes the polarised line profiles (if not already present), and computes a full Stokes formal solution.
    \item \texttt{prd\_redistribute} which performs a number of PRD subiterations, until either the maximum number of sub-iterations is performed, or the update size falls under a configurable tolerance.
    \item \texttt{stat\_equil} which computes the solution of the statistical equilibrium equations given the previously computed $\Gamma$ operator.
    \item \texttt{time\_dep\_update} which computes the solution of the kinetic equilibrium equations for one step of a provided duration.
    \item \texttt{nr\_post\_update} which computes the self-consistent electron density following Section \ref{Sec:EleDens}. In the case of statistical equilibrium this is called automatically, if the \texttt{Context} was initialised in charge conservation mode.
    \item \texttt{update\_deps} which updates various quantities such as the background opacity and line profiles to handle modifications to the model atmosphere (e.g. computing a finite-difference response function or changing timesteps in the case of a time-dependent simulation).
    \item \texttt{compute\_rays} which can compute a formal solution for one or multiple different viewing angles, optionally with full Stokes RT.
\end{itemize}

The \texttt{Context} and all other associated Cython components have been designed to support the Python \texttt{pickle} serialisation and deserialisation standard.
Therefore it is possible to save an entire context context to disk, and reload it and continue processing with only a few lines of code.
This will be discussed further in Section \ref{Sec:Parallelisation}.

\subsection{Parallelisation}\label{Sec:Parallelisation}

Two forms of parallelisation are supported by \Lw{}, one explicitly, and the other implicitly.
The \texttt{Context} object explicitly supports parallelisation of the formal solver over multiple threads of a single process by splitting the wavelengths over threads.
This approach also applies to PRD lines, for which the scattering integral can be computed in parallel.

When the aim is to process multiple atmospheres, for example in the case of a finite difference response function or a 1.5D atmosphere simulation, it is more efficient to use \Lw{} in a multi-process mode.
This can be done simply using e.g. \texttt{ProcessPoolExecutor}\footnote{\url{https://docs.python.org/3/library/concurrent.futures.html}} from the Python standard library for single computing nodes, or a Python MPI implementation for a multi-node cluster.
This method of computing is supported by the ability to \texttt{pickle} the \texttt{Context}, allowing the entire simulation state to be shipped between nodes in a single block if desired (although messages this large are taxing on process interconnects, and in many cases it is simpler to simply ship the data necessary to reconstruct the \texttt{Context} on a different node).

\subsection{Example}\label{Sec:Example}

The code in Listing \ref{Lst:CaComp} presents a simple script for running the comparison between the line profiles obtained for \CaLine{} with different electron densities in the FAL C atmosphere \citep{Fontenla1993}.

\begin{figure*}
    \centering
    \includegraphics[width=0.7\textwidth]{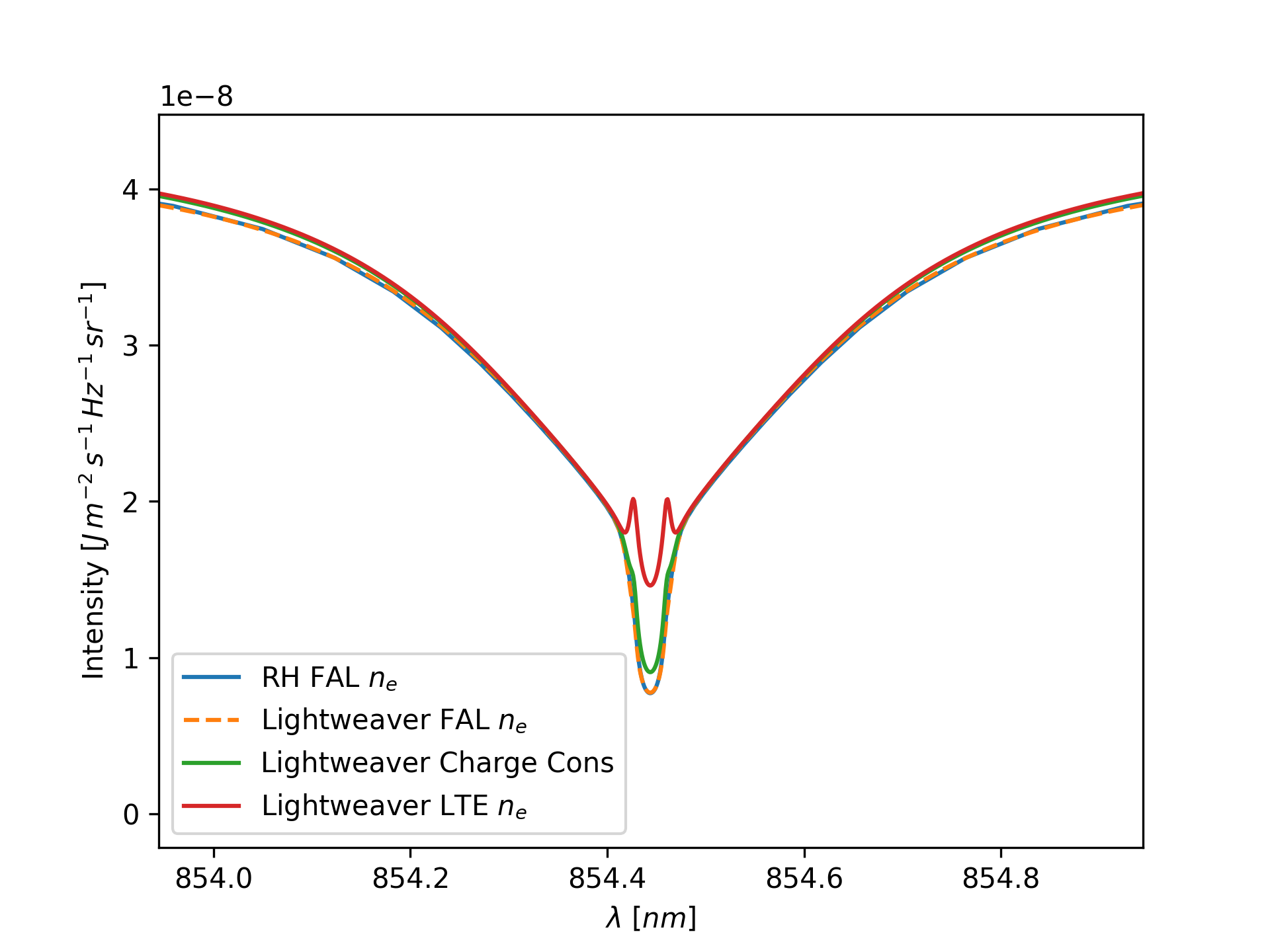}
    \caption{Comparison between RH and \Lw{} for the \CaLine{} line with different electron density solutions.}
    \label{Fig:RhComparison}
\end{figure*}

These are plotted against RH's solution for the electron density given in FAL C in Figure \ref{Fig:RhComparison}.

\begin{listing*}
\centering
\includegraphics{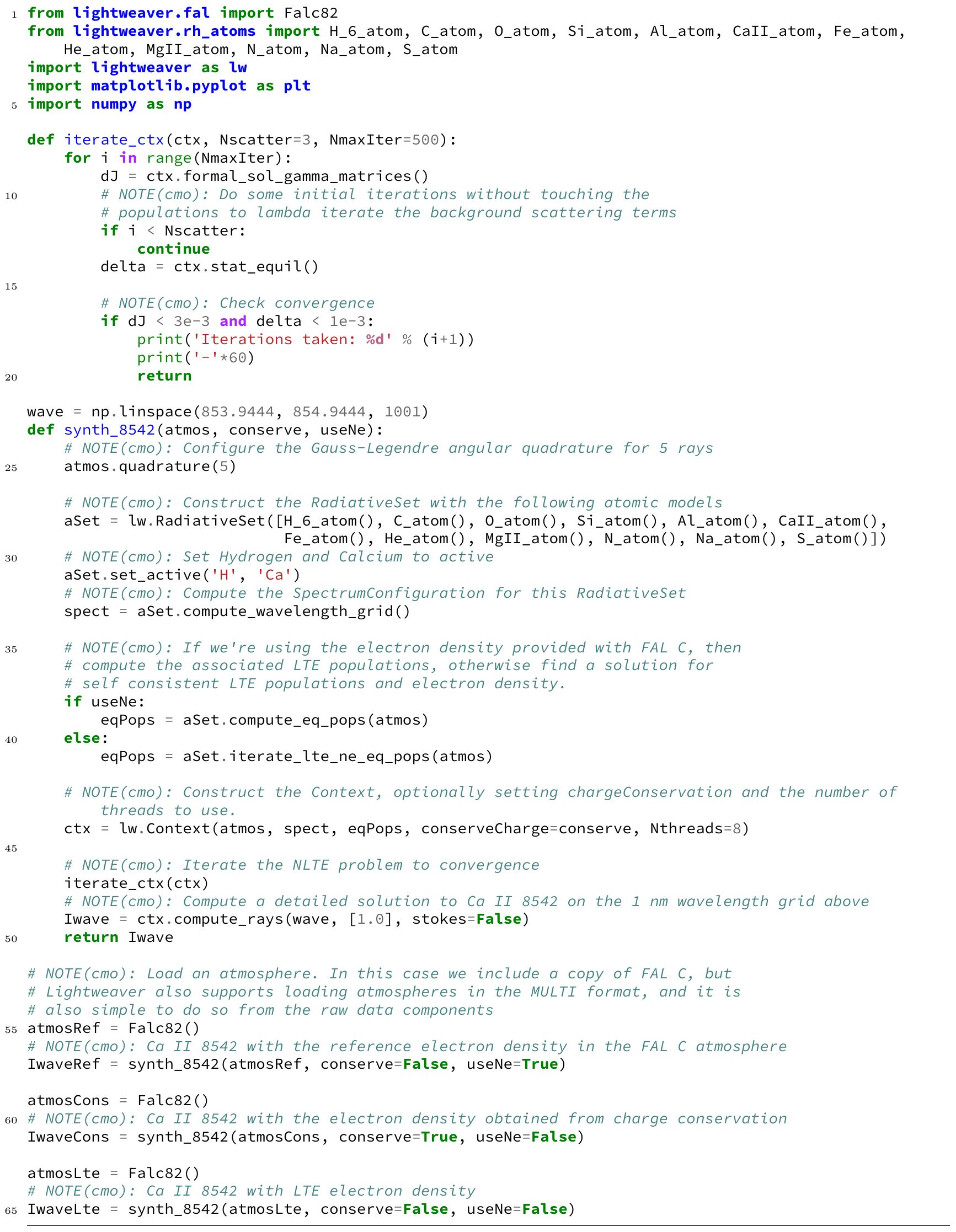}
\caption{Simple program comparing the results for \CaLine{} with different electron densities.}
\label{Lst:CaComp}
\end{listing*}

The agreement between RH and \Lw{} is extremely good when solving the same problem, as shown by the blue and dashed orange curves.
The red curve, using LTE electron density, shows the importance of the correct electron density for \CaLine{}.
The green curve, which is the solution computed with charge conservation from an LTE starting solution, approaches the reference electron density solution, and represents a self-consistent solution for electron density taking into account H and Ca in NLTE, however the final differences between the charge conserved solution and the reference solution are probably due to other elements, such as Fe, being treated in LTE.

\subsection{Advanced Example}\label{Sec:AdvancedExample}

In this section we present a more advanced example, first demonstrating the implementation of a different line profile (in this case Doppler) in a Ca\,\textsc{ii} model atom, and then using this modified model in a program that reprocesses output from the RADYN \citep{Carlsson1992,Allred2015} radiation-hydrodynamic code in a time-dependent fashion.

Listing~\ref{Lst:CaDop} demonstrates the modification of a 5 level with continuum Ca\,\textsc{ii} atom to use Doppler line profiles.
The \texttt{DopplerLine} class is first defined, with a new implementation of the \texttt{compute\_phi} method expected on an instance of \texttt{AtomicLine}.
It is then necessary to define the \texttt{NoOpBroadener} for the \texttt{LineBroadening} object provided to these lines, this class does nothing but provide comparison against itself and a \texttt{\_\_repr\_\_} method, allowing the model atom to be constructed from \texttt{repr} as discussed in Section~\ref{Sec:AtomicModels}.
Finally, the model atom is constructed as before, but using the newly defined \texttt{DopplerLine} class.
In this way model atoms can contain features such as different line profiles and collision rate parameterisations that are not known to the core \Lw{} package but remain compartmentalised in user code.

\begin{listing*}
\centering
\includegraphics{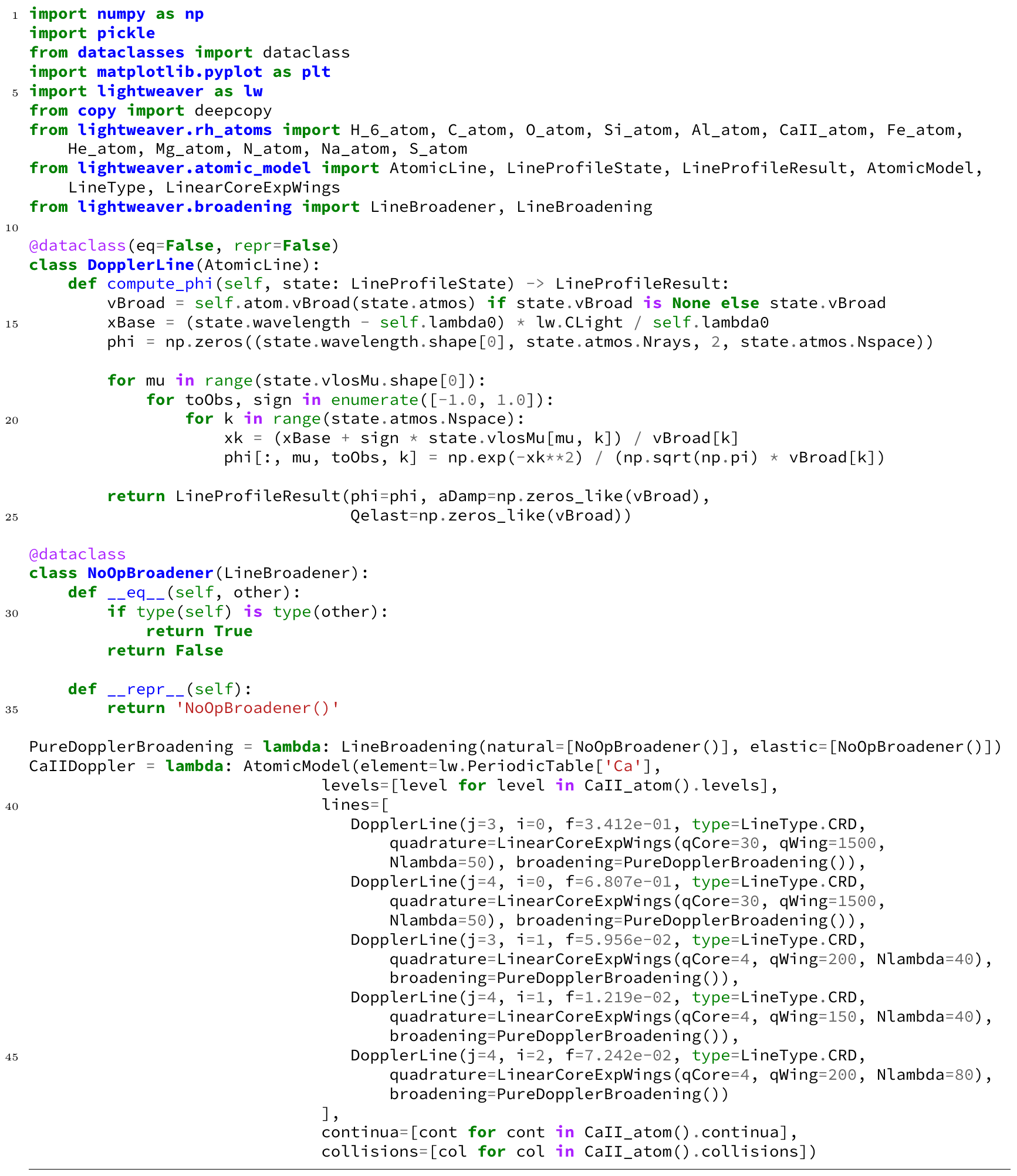}
\caption{Configuring a Ca\,\textsc{ii} model atom with Doppler profiles}
\label{Lst:CaDop}
\end{listing*}

Listing~\ref{Lst:CaTimeDep} shows the small amount of code needed to construct a specialised program for synthesising radiation from a pre-processed RADYN simulation (where the thermodynamic parameters of the atmosphere have been interpolated onto a fixed spatial grid) in a time-dependent fashion.
The simple method presented here ignores the advection of the populations by the plasma flows, but updated the calcium populations in a time-dependent fashion.
The hydrogen populations are loaded from the RADYN output and are used directly in the ``detailed static'' mode of operation.
Combining the Ca\,\textsc{ii} atom with Doppler line profiles from Listing~\ref{Lst:CaDop} and Listing~\ref{Lst:CaTimeDep} it is easy to perform this same synthesis twice, once with the traditional Voigt profiles and once with the Doppler profiles. The code present in these listings with the data file in the associated repository \citep{LightweaverTrRepository} are all that is needed to run this simulation.

First, the pre-processed data is loaded and an atmosphere object \texttt{atmos} is constructed from the initial timestep of the data. Several functions are then defined:
\begin{itemize}
    \item \texttt{construct\_context\_for} constructs a \texttt{Context} for an atmosphere and collection of model atoms, similarly to Listing~\ref{Lst:CaComp}.
    \item \texttt{initial\_stat\_eq} computes the statistical equilibrium solution using this context similarly to \texttt{iterate\_ctx} in Listing~\ref{Lst:CaComp}.
    \item \texttt{load\_step} loads the thermodynamic atmospheric properties, and hydrogen populations from the chosen timestep into the \texttt{Context}, before recomputing the line profiles and background opacities via \texttt{ctx.update\_deps}.
    \item \texttt{compute\_time\_dependent\_profiles} then uses the \texttt{load\_step} to load each timestep of the data present, solve the radiative transfer problem, and advance the calcium populations in time. It returns a \texttt{list} of the outgoing radiation from each timestep in the data.
\end{itemize}
Finally these functions are applied twice to construct two different simulations, one for each of the different calcium model atoms used.

A complete reprocessing of the RADYN simulation, considering the effects of advection of the atomic populations is substantially more complex, and outside the scope of the core \Lw{} framework. However, after implementing a suitable advection scheme the code presented here could easily be adapted.

\begin{listing*}
\centering
\includegraphics{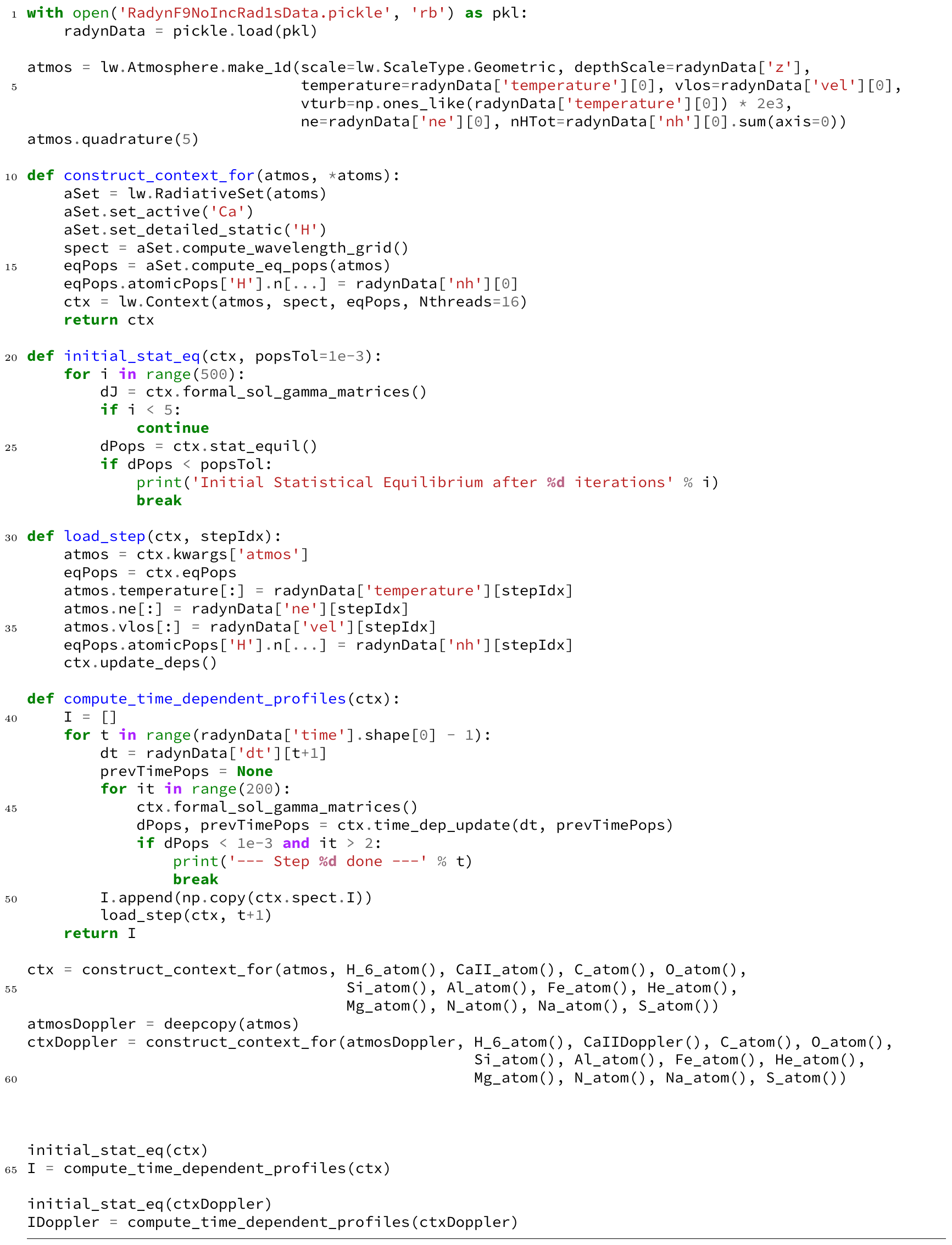}
\caption{Performing a basic time-dependent synthesis from RADYN simulation with and without Doppler line profiles in the Ca\,\textsc{ii} model atom}
\label{Lst:CaTimeDep}
\end{listing*}

\subsection{Performance Comparison}

Taking the FAL C example (with given electron density) presented in Section~\ref{Sec:Example}, for a 6-level model hydrogen atom and 6-level model calcium atom we will compare the performance of \Lw{} with RH.
These comparison tests were run on an Intel Xeon E3-1270 v3 (4 cores/8 logical threads, Haswell microarchitecture) with 1600 MHz DDR3 memory inside the Windows Subsystem for Linux environment in Microsoft Windows 10.0.18363.1016.
The compiled components of both codes were compiled with the GNU compiler collection 7.5.0, and \Lw{} was run using Python 3.8.2.

In both codes the atomic populations are initialised to LTE and Ng acceleration is disabled, to allow direct comparison of the iteration speed.
All tests were run five times, and the final result is the mean of these.
The variability from run-to-run is extremely low, so is not shown here.

\begin{table}
\centering
\begin{tabular}{ll}
\hline
Configuration                       & Time (s)      \\ \hline
RH wall time                        & 8.81                     \\
RH Setup \& Iteration Only          & 8.79                     \\
\Lw{} wall time                     & 11.05                    \\
\Lw{} Setup \& Iteration Only       & 8.47                     \\ \hline
\end{tabular}
\caption{Single-threaded comparison between RH and \Lw{} for a FAL C atmosphere with both H and Ca active.}
\label{Tab:ComparisonSt}
\end{table}

The single-threaded results are shown in Table~\ref{Tab:ComparisonSt}.
The difference the total wall time (real-world elapsed time) and the ``setup \& iteration only'' time is the time taken to load and correctly configure the model atmosphere.
The FAL C model used is defined on a column mass stratification.
Both RH and \Lw{} work in geometric height, so this stratification must first be converted.
This is a simple procedure with the total hydrogen density specified in the model atmosphere file.
To make the model atmosphere easy to manipulate on its own in \Lw{}, the continuum optical depth $\tau_{500}$ is also evaluated at the same time.
In RH this step takes place after the background opacities are computed, and these can be used directly (hence the very low cost of this step).
To improve flexibility in \Lw{}, this term uses background opacities obtained from the equation of state package discussed in Section~\ref{Sec:Eos}.
Many improvements could be made to the speed of this package by reimplementing its most numerically costly functions in a more performant language (or perhaps binding the pre-existing FORTRAN version to Python).
In practice, this one-off cost is rarely an issue as many models are now specified in terms of height, and \Lw{} does not require the calculation of column mass and continuum optical depth when a height stratified atmospheric definition is provided.

\begin{table}
\centering
\begin{tabular}{ll}
\hline
Configuration                       & Time (s)      \\ \hline
RH wall time                        & 7.37                     \\
RH Setup \& Iteration Only          & 7.35                     \\
\Lw{} wall time                     & 5.32                     \\
\Lw{} Setup \& Iteration Only       & 2.49                     \\ \hline
\end{tabular}
\caption{Multi-threaded comparison between RH and \Lw{} for a FAL C atmosphere with both H and Ca active.}
\label{Tab:ComparisonMt}
\end{table}

The codes were also compared when running on multiple threads.
In this case 8 threads were used in both codes, as this provided the fastest execution on this system.
These results are shown in Table~\ref{Tab:ComparisonMt}.
The constant cost of the $\tau_{500}$ conversion in \Lw{} can again be seen in these results.
\Lw{}'s threading model, that uses a thread-pool and lockless accumulation of the Gamma operator, provides significantly faster results at the cost of slightly higher memory consumption (one copy of the $\Gamma$ matrix per atom per thread, and one copy of the accumulation terms (see Appendix~\ref{App:MaliImpl}) per atom per thread).
Ignoring the aforementioned expensive one-off cost of computing $\tau_{500}$ for the model atmosphere via the EOS, RH achieves a 1.2x speedup from utilising multiple threads, whereas \Lw{} achieves a 3.4x speedup.
Accounting for the high cost of RH's threading model on Windows (where thread creation is very costly), this test was also run on a computer running CentOS 7, where a maximum speedup of 1.5x was recorded.

\section{Conclusions}

We have presented a brief overview of NLTE radiative transfer, and the methods used to solve associated problems employed by the \Lw{} framework.
We have also discussed the design of the framework, and hope that it will allow simpler experimentation with RT methods due to its ``factoring out''
of common operations into composable building blocks, and providing a single language approach to running and analysing simulations thanks to the extensive pre-existing set of scientific tools available in Python.
The nature of the framework allows programs for specialised tasks to be written far more easily than is possible in the traditional ``configuration file'' based monolithic code.
We are currently working on multi-dimensional extensions to the framework, to allow the synthesis of radiation from 2D and simple 3D atmospheres, but do not anticipate applying the advanced domain decomposition techniques of e.g. Multi3d \citep{Leenaarts2009} or PORTA \citep{Stepan2013}, however such an extension would be relatively simple thanks to the modularity of the codebase and simple serialisation of \texttt{Context} state.

\subsection*{Acknowledgments}

C.M.J.O. acknowledges support from the UK's Science and Technology Facilities Council (STFC) doctoral training grant ST/R504750/1 and is grateful for the financial aid of the STFC, CU Boulder, and the National Solar Observatory (NSO) for allowing the research trip to NSO where \Lw{} was designed.
The authors are also grateful to the reviewer for helpful comments and suggestions to improve the manuscript.

\Lw{} builds on the huge efforts of the scientific Python community and we acknowledge the NumPy \citep{Harris2020}, SciPy \citep{Virtanen2020}, Matplotlib \citep{Hunter2007}, Cython \citep{Behnel2011}, and Astropy \citep{Robitaille2013,Price-Whelan2018} packages.


\bibliography{Refs}{}
\bibliographystyle{aasjournal}

\appendix
\section{Line Broadening}\label{App:LineBroadening}

In the following we describe how the different broadening terms arise and are
implemented in \Lw{} as the basis for the standard Voigt line profile.
Natural broadening arises due to the finite lifetime of atomic states (described by $A_{ji}$) and the consequent variation in transitional energy due to the Heisenberg Uncertainty Principle.
The energy of an atomic transition is no longer perfectly defined, and instead takes the form of a Lorentzian distribution with damping coefficient $\Gamma_\mathrm{rad} = \sum_{i < j} A_{ji}$.
This description only accounts for broadening due to spontaneous emission, and a strong radiation field can modify this.
It may sometimes be useful to tune this parameter to account for observations, and lines that aren't present in a simplified model atom; $\Gamma_\mathrm{rad}$ is therefore a free parameter for each line in \Lw{}.

Doppler broadening is due to the random thermal motions of particles within the plasma.
In \Lw{}, we take the broadening velocity to be $v_\mathrm{broad} = \sqrt{2k_B T/W + v_\mathrm{turb}^2}$, where $W$ is the particle's mass, and $v_\mathrm{turb}$ is the microturbulent velocity specified for the atmosphere in question.
As Doppler broadening produces a Gaussian line profile, and radiative broadening produces a Lorentzian, the line profile due to both of these effects is the convolution of these, known as a Voigt profile.
Now, normalising with respect to the Doppler width
\begin{equation}
    \Delta \nu_D = \frac{v_\mathrm{broad} \nu_{ij}}{c},
\end{equation}
we have the absorption profile
\begin{equation}
    \varphi_{ij}(x) = \frac{H(a_\mathrm{damp}, x)}{\sqrt{\pi}},
\end{equation}
where
\begin{equation}
    x = \frac{\nu - \nu_{ij}}{\Delta \nu_D},
\end{equation}
and
\begin{equation}
    a_\mathrm{damp} = \frac{\Gamma}{4 \pi \Delta \nu_D}.
\end{equation}
$\Gamma$ is the sum of all ``typical'' broadenening terms (radiative, Stark, and van der Waals), and $H(a, x)$ is the Voigt function ($\int_{-\infty}^\infty H(a, x)\,dx = \sqrt{\pi}$).
The normalisation is such that
\begin{equation}
    \label{Eq:ProfileNorm}
    \frac{1}{4\pi} \oint \int \varphi_{ij}(x, \vec{d})\,dx\,d\Omega = 1
\end{equation}
when accounting for a directionally varying line profile.
The line profile $\varphi$ describes the photon absorption probability at a certain position in Doppler units.
To be dimensionally consistent with an integration over frequency we define the frequency dependent line profile $\phi_{ij}$ such that
\begin{equation}
\int_0^\infty \phi_{ij}\,d\nu=1
\end{equation}
i.e.
\begin{equation}
\phi_{ij} = \frac{\varphi_{ij}}{\Delta \nu_D}.
\end{equation}
Similarly to the line absorption profile, the properties of a continuum are controlled by the atomic cross-section at a given frequency.
The cross-sections can be described in two different ways in \Lw{}: either as a \textit{hydrogenic} continuum, in which case the cross-section falls off as $1/\nu^3$ with increasing frequency from the continuum edge, or as a tabulated cross-section, whereby the cross-section is provided at different wavelengths, and then interpolated between.

Note that the quantity stored in the variable \texttt{phi} in the code is in fact $\varphi / v_{\mathrm{broad}}$, as this simplifies construction of some of the expressions.
We denote this term $\phi_{\mathrm{num}}$, which is used in Section \ref{App:MaliImpl} when describing the calculation of emissivity and opacity.

\section{MALI Numerical Implementation}\label{App:MaliImpl}

In the following the integration and accumulation terms used in the implementation of the MALI method (Section~\ref{Sec:Mali}) are presented.

As discussed in Section~\ref{Sec:MaliImpl} a common wavelength grid is first computed from which the final individual wavelength grid for each transition is extracted.
For each transition with its final individual discrete wavelength grid (denoted $\lambda$) we define the integration weights
\begin{equation}
    w_{\lambda_a} =
    \begin{cases}
        0.5 \left( \lambda_{a + 1} - \lambda_{a} \right) D_{ij}, & a = a_{\min}  \\
        0.5 \left( \lambda_{a} - \lambda_{a + 1} \right) D_{ij}, & a = a_{\max} \\
        0.5 \left( \lambda_{a+1} - \lambda_{a-1} \right) D_{ij}, & \textrm{otherwise},
    \end{cases}
\end{equation}
with
\begin{equation}
    D_{ij} =
    \begin{cases}
        c / \lambda_{ij} & \textrm{bound-bound} \\
        1 & \textrm{bound-free}.
    \end{cases}
\end{equation}

For lines, this can be used to compute the line-profile normalisation factor
\begin{equation}
    w_{\phi, k} = \left(\sum_{i, \mu} \phi_{\mathrm{num}, \lambda_i, \vec{d}} w_{\lambda_i} w_{\mu} \right)^{-1}
\end{equation}
that ensures that \eqref{Eq:ProfileNorm} holds for each discretised point in the atmosphere $k$. $\phi_{\mathrm{num}}$ is defined in Appendix~\ref{App:LineBroadening}.
This line-profile normalisation factor is essential in ensuring that $\Gamma$ is correctly scaled, especially for more sparsely sampled lines.
The integration weights for the terms contributing to $\Gamma^R$ for each transition $ij$ at each depth are then
\begin{equation}
    w_{\Gamma, ij, \lambda_a} =
    \begin{cases}
        w_{\lambda_a} w_\phi 4\pi / (h c), & \textrm{bound-bound} \\
        w_{\lambda_a} 4\pi / (h \lambda_a), & \textrm{bound-free}
    \end{cases}
\end{equation}
these terms may not be immediately evident upon comparison with \eqref{Eq:GammaR}; the differences arise from ensuring that all terms are integrated in frequency, despite the discretisation being performed in wavelength.

From the previous discussion of line profiles in Appendix~\ref{App:LineBroadening}, we have
\begin{equation}
    \frac{h\nu}{4\pi}\phi_{ij} = \frac{hc}{4\pi}\phi_{\mathrm{num},\,{ij}}
\end{equation}
under the reasonable approximation that $\nu=\nu_{ij}$ over the integration range for a spectral line.
This latter formulation is used when evaluating $U$ and $V$ in \Lw{}.
Additionally, we follow \citet{Uitenbroek2001} and define
\begin{equation}
    g_{ij} =
    \begin{cases}
        g_i / g_j \rho_{ij} = B_{ji} / B_{ij} \rho_{ij} & \textrm{bound-bound} \\
        n_i^* / n_j^* \exp{\left( -\frac{h\nu}{k_B T} \right)} & \textrm{bound-free},
    \end{cases}
\end{equation}
such that
\begin{equation}
    V_{ji} = g_{ij} V_{ij},
\end{equation}
and
\begin{equation}
    U_{ji} = \frac{2h\nu^3}{c^2} V_{ji},
\end{equation}
which can be expressed as $U_{ji} = A_{ji} / B_{ji}  V_{ji}$ for bound-bound transitions.
With these expressions the $U$ and $V$ terms are very efficient to evaluate in a vectorised manner, and we therefore choose to not cache them. If one were writing an entirely CRD code, these terms can be computed once per transition and stored, which could be a valuable optimisation in some cases, at the expense of computer memory.

During the accumulation of emissivities and opacities at each wavelength and direction for the formal solver we also accumulate the emissivity per atom
\begin{equation}\label{Eq:EtaAccum}
    \mathcal{H} = \sum_{i,j}\eta^\dagger_{ij},
\end{equation}
total $U_{ji}$ per level $j$
\begin{equation}
    \mathcal{U}_j = \sum_i U^\dagger_{ji},
\end{equation}
and effective self-opacity in for each level $l$
\begin{equation}\label{Eq:ChiAccum}
    \mathcal{X}_l = \sum_{j > l} \chi^\dagger_{lj} - \sum_{i < l} \chi^\dagger_{il}.
\end{equation}

If no lines are associated with a wavelength point, then all the sources of emissivity and opacity are direction independent, and these accumulations can only be performed once for all of the directional formal solutions.
As the accumulation of terms into $\Gamma^R$ can be done after the formal solution for each direction is performed, the storage of these terms does not increase with the number of directions.
If one adopts the ``same-transition'' preconditioning approach of \citep{Rybicki1992} then none of these accumulation arrays are needed.
This could be advantageous in the implementation of higher dimensional schemes, as in most cases there appears to be no difference in convergence speed between the two methods.

The formal solver then provides the values of $I^\dagger(\nu, \vec{d})$ and $\Psi^*_{\nu, \vec{d}}$, for all depths, one direction at a time.
The per-atom $I^\mathrm{eff}$ term of \eqref{Eq:Ieff} for this direction can then simply be computed as
\begin{equation}
    I^\mathrm{eff}_{\nu, \vec{d}} = I^\dagger(\nu, \vec{d}) - \Psi^*_{\nu,\vec{d}} \mathcal{H},
\end{equation}
where the final term is a simple scalar multiplication, under the assumption that $\Psi^*$ is simply the diagonal of the true $\Psi$ operator (which is currently the case in \Lw{}).

With these definitions the integration of the off-diagonal entries of $\Gamma^R$ at each spatial point, for each active atom, is performed by looping over each contributing transition $ij$ such that
\begin{align}
    \Gamma_{ij}^R &= \sum_{a, \mu} \left[ w_\mu w_{\Gamma, ij, \lambda_a} \left( U_{ji} + V_{ji} I^\mathrm{eff} - \mathcal{X}_i \Psi^* \mathcal{U}_j \right) \right], \\
    \Gamma_{ji}^R &= \sum_{a, \mu} \left[ w_\mu w_{\Gamma, ij, \lambda_a} \left( V_{ij}I^\mathrm{eff} - \mathcal{X}_j \Psi^* \mathcal{U}_i \right) \right].
\end{align}

The radiative rates are computed similarly
\begin{align}
    R_{ij} &= \sum_{a, \mu} \left[ w_\mu w_{\Gamma, ij, \lambda_a} I^\dagger V_{ij} \right], \\
    R_{ji} &= \sum_{a, \mu} \left[ w_\mu w_{\Gamma, ij, \lambda_a} \left( U_{ji} + I^\dagger V_{ij} \right) \right].
\end{align}

\section{PRD Implementation}\label{App:CompletePrdImpl}

In the following we describe the terms needed to apply PRD to a spectral line and their implementation in \Lw{}.
For the previous definition of $\rho_{ij}$ from \eqref{Eq:RhoDef}, under the assumptions of a line with an infinitely sharp lower level and broadened upper level, and the validity of PRD being in the atomic frame being approximated by PRD in the observer's frame \citep{Uitenbroek2001}, following \citet{Hubeny2014} we then have
\begin{align}
\begin{split}
    \rho_{ij}(\nu, \vec{d}) = 1 + &\gamma\frac{\sum_{l < j}n_j B_{lj}}{n_j P_j} \oint\frac{1}{4\pi}\int I(\nu^\prime, \vec{d}^\prime) \\ &\cdot \left[ \frac{R^{II}_{lji}(\nu^\prime, \vec{d}^\prime; \nu, \vec{d})}{\phi_{ij}(\nu, \vec{d})} - \phi_{lj}(\nu^\prime, \vec{d}) \right]\,d\nu^\prime\,d\Omega^\prime,
\end{split}
\end{align}

where $R^{II}$ is the generalised redistibution function for transitions of this kind \citep{Hubeny1982}, and $\gamma$ is the branching ratio, or coherency fraction.
The summation over $l$ and $lji$ subscript on $R^{II}$ describe the scattering process.
When ignoring cross-redistribution (Raman scattering), we have $l=i$, and the summation is replaced by a single term, as we are only considering resonance scattering within the line $i\rightarrow j$.

The coherency fraction $\gamma$ describes the normalised probability of a photons being re-emitted from the same sublevel of energy level $j$ before an elastic collision that will redistribute it across sublevels, provided that it is re-emitted at all.
This is then given by
\begin{equation}
    \gamma = \frac{P_j}{P_j + Q_j},
\end{equation}
where $P_j$ is the total rate of transitions out of level $j$ (depopulation rate), and $Q_j$ is the total rate of elastic collisions affecting this level.

Defining $g_{II}(\nu, \nu^\prime) = R^{II}(\nu, \nu^\prime)/\phi_{ij}(\nu^\prime)$, which is normalised such that
\begin{equation}
    \label{Eq:gIINorm}
    \frac{1}{4\pi}\oint\int g_{II}(\nu, \nu^\prime) \,d\nu^\prime\,d\Omega= 1,
\end{equation}
as per \citet{Gouttebroze1986} and \citet{Uitenbroek1989} wherein fast approximations to this function are derived, and ignoring cross-redistribution we then have
\begin{align}
\begin{split}
    \label{Eq:RhoPrd}
    \rho_{ij}(\nu, \vec{d}) = 1 + & \gamma \frac{n_i B_{ij}}{n_j P_j} \oint \frac{1}{4\pi} \int I(\nu^\prime, \vec{d}^\prime) \\ & \cdot\left[ g_{II}(\nu, \nu^\prime) - \phi_{ij}(\nu^\prime, \vec{d}) \right]\,d\Omega^\prime\,d\nu^\prime.
\end{split}
\end{align}

Ignoring bulk plasma flows, the integrals over angle and frequency can then be split, providing an angle-averaged form of $\rho$ that is much easier to compute, given by
\begin{equation}
    \label{Eq:RhoPrdAa}
    \rho_{ij}(\nu) = 1 + \gamma \frac{n_i B_{ij}}{n_j P_j} \left( \int g_{II}(\nu, \nu^\prime) J(\nu^\prime)\,d\nu^\prime - \bar{J}_{ij} \right),
\end{equation}
where
\begin{equation}
\bar{J}_{ij} = \frac{1}{4\pi} \oint \int I(\nu, \vec{d}) \phi(\nu, \vec{d})\,d\nu\,d\Omega = \frac{R_{ij}}{B_{ij}}
\end{equation}
is the frequency-integrated mean intensity across the transition.

The numerical implementation of angle-averaged PRD simply follows the method of \citet{Uitenbroek2001}, which is briefly summarised below.
The redistribution function is often very sharply peaked, an accurate evaluation of the scattering integral therefore requires much finer sampling of the wavelength grid than that which is required the evaluate the terms in $\Gamma^R$.
However for each wavelength in a line's grid there is only a small surrounding region for which $g_{II}$ is non-zero.
$J(\nu)$ is then interpolated onto a fine grid over this region, over which \eqref{Eq:RhoPrd} is trivially implemented by an application of Simpson's rule.
It is essential that the scattering integral component of \eqref{Eq:RhoPrd} be normalised as per \eqref{Eq:gIINorm} to avoid the addition or destruction of photons in the transition.
The term
\begin{equation}
    \int g_{II}(\nu, \nu^\prime)J(\nu^\prime) d\nu^\prime
\end{equation}
is then implemented as
\begin{equation}
    \frac{\sum_i g_{II}(\nu, \nu^\prime_i) J(\nu^\prime_i) \delta \nu^\prime_i}{\sum_i g_{II}(\nu, \nu^\prime_i) \delta \nu^\prime_i},
\end{equation}
where $\delta \nu^\prime_i$ are the integration weights over this fine grid.

The iterative method currently employed consists of performing a formal solution over the wavelengths where PRD lines are active, and updating $\rho$ using this method \textit{whilst maintaining the populations fixed}.
When solving a PRD problem, a number of these sub-iterations to update $\rho$ (commonly 3) are interleaved between every complete formal solution and population update.

For the hybrid PRD case of \citet{Leenaarts2012} used when plasma flows exceed the thermal Doppler velocity, $J(\nu)$ in \eqref{Eq:RhoPrdAa} is then replaced by $J_\mathrm{rest}(\nu)$, the mean intensity in the atom's rest frame.
This approximation is much faster to evaluate than the full angle-dependent case, as the accumulation of $J_\mathrm{rest}(\nu)$ can be done during the formal solution, using a linear interpolation off the Doppler-shifted frequency grid.
$\rho$ can be linearly interpolated from the atomic rest frame during the calculation of the $U$ and $V$ terms, or into a directionally dependent array at the end of each PRD sub-iteration.
Currently \Lw{} does the former of these.

To ensure that $J_\mathrm{rest}$ is accumulated correctly during the PRD sub-iterations, we no longer simply perform formal solutions over wavelengths where PRD lines are present, but also over wavelengths that when shifted back to the rest frame contribute to $J_\mathrm{rest}$ in these regions.
We have found that this modification, that is not present in RH1.5D \citep{Pereira2015}, can dramatically aid convergence in atmospheres with high velocity shifts.

\section{Self-Consistent Newton-Raphson Electron Density Iteration}\label{App:EleDens}

From Section \ref{Sec:Mali}, the equations of statistical equilibrium (ESE) are given by \eqref{Eq:StatEq}.
Let us write this system for a level $i$ of a species $s$ as
\begin{equation}
    \label{Eq:EseFn}
    F_{s, i}(\vec{n}_s, n_e) = \sum_{j\neq i} n_j P_{ji}(\vec{n}_s, n_e) - n_i\sum_{j\neq i} P_{ij}(\vec{n}_s, n_e) = 0
\end{equation}
With a fixed electron density the preconditioned linear formulation of these equations for a species $s$ can be written
\begin{equation}
    \Gamma_s \vec{n}_s = \vec{0}.
\end{equation}
We start by obtaining the solution to this linear system that, in the following, will be denoted $\widetilde{\vec{n}_s}$.
The previous values of these populations, i.e. the ones at which $\Gamma_s$ was evaluated, will once again be denoted with $\dagger$.

The principle of Newton-Raphson iteration is to compute $F(x_0 + \delta x) = 0$, which can be written as $F(x_0) + J\delta x = 0$, with $J$ the Jacobian of $F$ evaluated at $x_0$, and $\delta x$ some small correction to $x_0$.
This can be rearranged to $-J\delta x = F(x_0)$.
Applying this to technique \eqref{Eq:EseFn} and explanding to first order we have
\begin{align}
\begin{split}
    \label{Eq:LinNr}
    &- \sum_j \left( \left.\frac{\partial F_{s,i}(\vec{n}_s, n_e)}{\partial n_j}\right\rvert_{(\widetilde{\vec{n}_s}, n_e^\dagger)} \delta n_j \right)\\
    &\hspace{2.5em}- \left.\frac{\partial F_{s,i}(\vec{n}_s, n_e)}{\partial n_e}\right\rvert_{(\widetilde{\vec{n}_s}, n_e^\dagger)} \delta n_e \\ & = F_{s, i}(\widetilde{\vec{n}_s}, n_e^\dagger),
\end{split}
\end{align}
where $\delta n_j$ and $\delta n_e$ indicate the corrections to these populations necessary to render them self-consistent.
This expression can be written for each level of each active species.
Looking more closely at each term we have
\begin{equation}
    \label{Eq:EseDerivN}
\left.\frac{\partial F_{s,i}(\vec{n}_s, n_e)}{\partial n_j}\right\rvert_{(\widetilde{\vec{n}_s}, n_e^\dagger)} = \Gamma_{s, ij},
\end{equation}
and
\begin{align}
\begin{split}
    \label{Eq:EseDerivNe}
    &\left.\frac{\partial F_{s,i}(\vec{n}_s, n_e)}{\partial n_e}\right\rvert_{(\widetilde{\vec{n}_s}, n_e^\dagger)} =\\
    &\hspace{2em}\sum_j\left(\left.\frac{\partial \Gamma_{s, ij}^C}{\partial n_e}\right\rvert_{(\widetilde{\vec{n}_s}, n_e^\dagger)} \widetilde{n_j} \right)\\
    &\hspace{2em}+\sum_j
    \begin{cases}
    \Gamma_{s, ij}^R \widetilde{n_j} / n_e^\dagger , &i\rightarrow j\quad \textrm{bound-free}\\
    0, &\textrm{otherwise}
    \end{cases},
\end{split}
\end{align}
where the term involving $\Gamma^C$ depends on the exact form of the collisional rates.
Due to the number of collisional rate options available in \Lw{} this is evaluated through finite differences, which remains relatively efficient due to the local nature of this term.

As can be seen from \eqref{Eq:EseDerivN} and \eqref{Eq:EseDerivNe}, all terms are linear in $\delta n$ and $\delta n_e$, however additional constraints are needed to close this system: a constraint on the total population of each species $s$, and a constraint on charge neutrality.
In total, this forms a system $\sum_s N_{\mathrm{level}, s} + 1$ equations, where $N_{\mathrm{level}, s}$ is the number of levels treated in detail for species $s$, and the summmation is performed over all active species.
This system can therefore be written at each point in the atmosphere as a block diagonal matrix, with each block being $N_{\mathrm{level}, s} \times N_{\mathrm{level}, s}$, with a final row and column due to the electron density terms and charge conservation equation that couple all blocks.
The block terms are simply $-\Gamma_s$ for each species, and the final column for level $j$ of species $s$ is the additive inverse of the right-hand side of \eqref{Eq:EseDerivNe}.
In our implementation the population conservation equation is
\begin{equation}
    \sum_j \delta n_j = n_\mathrm{total} - \sum_j \widetilde{n_j},
\end{equation}
where $n_\mathrm{total}$ is the total population of the species (derived from abundance and hydrogen density, or mass density).
The left-hand side amounts to placing a block row of ones in the Jacobian for all levels in the species.
In our case we replace the last equation for each species with this population conservation equation to avoid degeneracies.

The charge conservation equation is given by
\begin{equation}
    \delta n_e - \sum_s \sum_j \mathion_s(j) \delta n_{s, j} = n_{e, \mathrm{bg}} + \sum_s \sum_j \mathion_s(j) - n_e^\dagger,
\end{equation}
where $\mathion_s(j)$ is the ionisation level of the $j$-th level of species $s$, and $n_{e, \mathrm{bg}}$ is the electron density due to background species whose populations are not otherwise taken into account during this iteration.
The left hand side is inserted into the final row of the Jacobian.

The right-hand side vector for the Newton-Raphson procedure, where not specified by the constraint equations, is given by $\Gamma_s \widetilde{\vec{n}_s}$.
This system can be solved as a typical matrix-vector system to obtain the corrections.
Finally, the populations are corrected as $n = \widetilde{n} + \delta n$ and $n_e = n_e^\dagger + \delta n_e$, and all LTE populations must be updated for consistency with the new electron density.
As hydrogen is by far the dominant contributor of electrons, we optionally allow the above to only operate on hydrogen, and count all other species as background for the purpose of updating $n_e$, and in some cases this may be more stable.

The time-dependent case then follows a similar derivation to the statistical equilibrium case.
Here we start from the $\theta$-method of \eqref{Eq:TimeDepSystem} and similarly to \eqref{Eq:EseFn} define
\begin{align}
\begin{split}
    \label{Eq:TimeDepNrFn}
    G_{s,i}(\vec{n}^{t+1}_s, n_e) = n^{t+1}_{s,i} - &\theta \Delta t F_{s,i}(\vec{n}^{t+1}_s, n_e) \\- &(1-\theta)\Delta t \Gamma^t_s \vec{n}^t_s - n^t_{s,i} = 0.
\end{split}
\end{align}

Similarly to \eqref{Eq:EseDerivN} we then have
\begin{equation}
    \left.\frac{\partial G_{s,i}(\vec{n}^{t+1}_s, n_e)}{\partial n_j}\right\rvert_{(\widetilde{\vec{n}^{t+1}_s}, n_e^\dagger)} = \delta_{ij} - \theta \Delta t \Gamma_{s,ij},
\end{equation}
where $\delta_{ij}$ is the Kronecker delta, and then similarly to \eqref{Eq:EseDerivNe}
\begin{equation}
    \left.\frac{\partial G_{s,i}(\vec{n}^{t+1}_s, n_e)}{\partial n_e}\right\rvert_{(\widetilde{\vec{n}^{t+1}_s}, n_e^\dagger)} = -\theta \Delta t
    \left.\frac{\partial F_{s,i}(\vec{n}^{t+1}_s, n_e)}{\partial n_e}\right\rvert_{(\widetilde{\vec{n}^{t+1}_s}, n_e^\dagger)}.
\end{equation}

The Jacobian matrix is constructed in the same way as the statistical equilibrium case, but using the derivatives of $G$.
The right-hand side vector of the Newton-Raphson iteration procedure is then given by \eqref{Eq:TimeDepNrFn}, where the superscript $t$ terms are known from the start of the timestep and need to be stored for use in this procedure.
The constraint equations remain the same as in the time-independent case.
As in the case of time-dependent population updates, \Lw{} currently only supports the $\theta=1$ case, but the ground-work is present for supporting other $\theta$.



\end{document}